\input{aipcheck}

\documentclass[
    ,final            
  ]
  {aipproc}

\layoutstyle{6x9}

\newcommand{\Slash}[1]{\ooalign{\hfil/\hfil\crcr$#1$}}

\begin{document}

\title{Resonance interpretation of the bump structure in the $\gamma p \to K^+ \Lambda(1520)$ differential cross section}


\author{Ju-Jun Xie}{
  address={Instituto de F\'\i sica Corpuscular (IFIC), Centro Mixto
CSIC-Universidad de Valencia, Institutos de Investigaci\'on de
Paterna, Aptd. 22085, E-46071 Valencia, Spain},
 altaddress={Department of Physics, Zhengzhou University, Zhengzhou, Henan 450001, China}}
\author{J. Nieves}{
  address={Instituto de F\'\i sica Corpuscular (IFIC), Centro Mixto
CSIC-Universidad de Valencia, Institutos de Investigaci\'on de
Paterna, Aptd. 22085, E-46071 Valencia, Spain} }


\begin{abstract}

We investigate the $\Lambda(1520)$ photoproduction in the
$\vec{\gamma} p \to K^+ \Lambda(1520)$ reaction within the effective
Lagrangian method near threshold. In addition to the "background"
contributions from the contact, $t-$channel $K$ exchange, and
$s-$channel nucleon pole terms, which were already considered in
previous works, the contribution from the nucleon resonance
$N^*(2080)$ (spin-parity $J^P = 3/2^-$) is also considered. We show
that the inclusion of the nucleon resonance $N^*(2080)$ leads to a
fairly good description of the new LEPS differential cross-section
data, and that these measurements can be used to determine some of
the properties of this latter resonance.

\end{abstract}

\maketitle


\section{Introduction}

The $\Lambda(1520)$ ($\equiv \Lambda^*$) photoproduction in the
$\gamma p \to K^+ \Lambda^* $ reaction is an interesting tool to
gain a deeper understanding of the interaction among strange hadrons
and also on the nature of baryon resonances. Recently, this reaction
has been examined at photon energies below $2.4$ GeV in the SPring-8
LEPS experiment~\cite{leps}. For an invariant $\gamma p$ mass $W
\simeq 2.11$ GeV, the experiment has reported a new bump structure
in the differential cross section at forward $K^+$ angles, which
might hint to a sizeable contribution from nucleon resonances in the
$s-$channel.

Different dominant mechanisms~\cite{titovprc7274,nam,toki} have been
proposed to describe the high-energy results from
LAMP2~\cite{lamp2}. However, all these models fail to describe the
forward bump structure in the new LEPS data. Thus, we reanalyze the
$\vec{\gamma} p \to K^+ \Lambda^* $ reaction within the effective
Lagrangian method. In addition to the "background" contributions
from the contact, $t-$channel $K$ exchange, and $s-$channel nucleon
pole terms, which were already considered in previous works, we have
also studied possible contributions from nucleon resonances.
Unfortunately, the information about nucleon resonances in the
relevant energy region ($\sim 2.1$) GeV is scarce~\cite{pdg2008},
and we should rely on theoretical schemes, such that of
Ref.~\cite{simonprd58} based in a quark model (QM) description of
hadrons. Among the possible nucleon resonances, we have finally
considered only the two-star $D-$wave $J^P=3/2^-$ $N^*(2080)$
($\equiv N^*$) one. Although the $N^*(2080)$ resonance is listed in
the Particle Data Group (PDG) book, the evidence of its existence is
poor or only fair and further work is required to verify its
existence and to determine its properties. In this respect, we show
here how the recent LEPS measurements could be used to determine
some of the properties of this resonance.

\section{Formalism} \label{sec:formalism}

\begin{center}
\begin{figure}[htdp]
\vspace{0.2cm}
\includegraphics[scale=0.4]{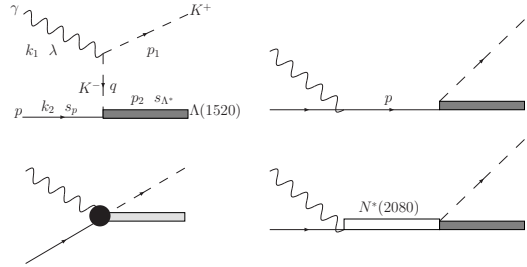}
\caption{Model for the $\gamma p \to \Lambda^* K^+$ reaction.}
\label{diagram}
\end{figure}
\end{center}

The basic tree level Feynman diagrams for the $\vec{\gamma} p \to
K^+ \Lambda^* $ reaction are depicted in the Fig.~\ref{diagram}. To
compute the contributions of each of these terms, we use the
interaction Lagrangian densities of Refs.~\cite{nam,toki},
\begin{eqnarray}
\mathcal{L}_{\gamma KK} &=& -ie (K^- \partial^{\mu} K^+ -
K^+\partial^{\mu}  K^- ) A_{\mu}, \\ \label{eq:gamakk}
\mathcal{L}_{Kp\Lambda^*} &=&
\frac{g_{KN\Lambda^*}}{m_{K}}\bar{\Lambda}^{*\mu} (\partial_{\mu}
K^-){\gamma}_{5}p\,+{\rm h.c.},  \label{eq:knl} \\
\mathcal{L}_{\gamma pp} &=&
-e\bar{p}\left(\Slash{A}-\frac{\kappa_{p}}{2M_{N}}
\sigma_{\mu\nu}(\partial^{\nu}A^{\mu})\right) p + {\rm h.c.}, \\
\mathcal{L}_{\gamma Kp\Lambda^*} &=&
-ie\frac{g_{KN\Lambda^*}}{m_{K}}\bar{\Lambda}^{*\mu} A_{\mu}
K^-{\gamma}_{5}p\,+{\rm h.c.}, \\ \label{eq:eqfin}
\mathcal{L}_{\gamma N N^*} &=& \frac{ie f_1}{2m_{N}} \bar{N}^*_{\mu}
\gamma_{\nu} F^{\mu \nu} N \, - \frac{e
f_2}{(2m_{N})^2}\bar{N}^*_{\mu} F^{\mu \nu} \partial_{\nu} N\,+{\rm
h.c.}, \\ \label{eq:eqgamaN} \mathcal{L}_{K \Lambda^* N^*} &=&
\frac{g_1}{m_{K}} \bar{\Lambda}^*_{\mu} \gamma_{5} \gamma_{\alpha}
(\partial^{\alpha} K) N^{* \mu} \, +  \frac{i g_2}{m_{K}^2}
\bar{\Lambda}^*_{\mu} \gamma_5 \left (\partial^{\mu} \partial_{\nu}
K\right)  N^{*\nu} \,+{\rm h.c.}, \label{eq:eqknstar}
\end{eqnarray}
where $e=\sqrt{4\pi\alpha} > 0$ ($\alpha = 1/137.036$ is the
fine-structure constant), $\kappa_p=1.79$, $A_\mu$ and $F_{\mu \nu} =
\partial_{\mu}A_{\nu}-\partial_{\nu}A_{\mu}$ are the proton charge and
magnetic moment, and the photon field and electromagnetic field
tensor, respectively.

With the effective interaction Lagrangian densities given above, we
can easily construct the invariant scattering amplitudes (see
Ref.~\cite{xiejuan} for its detailed expression). Moreover, we also
include the form factors, respecting gauge invariance as follows
\begin{eqnarray}
f_{\rm i} &=&\frac{\Lambda^4_i}{\Lambda^4_i+(q_{\rm i}^2-M_{\rm
i}^2)^2},
\quad {\rm i}={\rm s,t, R} \\
f_{\rm c} &=& f_{\rm s}+f_{\rm t}-f_{\rm s}f_{\rm t}, \quad {\rm
and} \left\{\begin{array}{l}                               q_{\rm
  s}^2=q_{\rm R}^2=s,\, q_{\rm t}^2= q^2, \cr
M_{\rm s} = M_N, \cr
 M_{\rm R} = M_{N^*}, \cr
M_{\rm t} = m_K.
\end{array}\right. \label{F1}
\end{eqnarray}
We will consider different cut-off values for the background and
resonant terms, i.e.  $\Lambda_s=\Lambda_t \ne \Lambda_R$.

\section{Numerical results and discussion} \label{sec:results}
The differential cross section, in the center of mass frame (C.M.),
and for a polarized photon beam reads,
\begin{eqnarray}
\hspace{-0.2cm}\frac{d\sigma}{d\Omega}\Big|_{\rm C.M.} &=&
\frac{|\vec{k}_1^{\rm \,\, C.M.}||\vec{p}_1^{\rm
\,\,C.M.}|}{4\pi^2}\frac{M_N M_{\Lambda^*}}{(s-M_N^2)^2} \, \Big
(\frac{1}{2}\sum_{s_p, s_\Lambda^*} |T|^2 \Big )
 \label{eqdcs}
\end{eqnarray}
where $\vec{k}_1^{\rm \,\, C.M.}$ and $\vec{p}_1^{\rm \,\, C.M.}$
are the photon and $K^+$ meson c.m. three-momenta.

We perform eight parameter ($ef_1$, $ef_2$, $g_1$,$g_2$,
$\Lambda_s=\Lambda_t$, $\Lambda_R$, $M_{N^*}$ and $\Gamma_{N^*}$ )
$\chi^2-$fits to the LEPS $d\sigma/d(\cos\theta_{\rm C.M.})$ data at
forward angles. There is a total of 59 data points. These data
correspond to forward $K^+$ angles and are given for four intervals
of $\cos\theta_{\rm C.M.}$ ranging from 1 down to 0.6. To compute
the cross sections in each interval we always use the corresponding
mean value of $\cos\theta_{\rm C.M.}$. The fitted parameters are
$ef_1 = 0.177 \pm 0.023$, $ef_2 = -0.082 \pm 0.023$, $g_1 = 1.4 \pm
0.3$, $g_2 = 5.5 \pm 1.8$, $\Lambda_s=\Lambda_t= 604 \pm 2$ (MeV),
$\Lambda_R=909 \pm 55$ (MeV), $M_{N^*}=2115 \pm 8$ (MeV), and
$\Gamma_{N^*}=254 \pm 24$ (MeV). We show the results in Figure 2.

\begin{center}
\begin{figure}[htdp]
\includegraphics[scale=0.6]{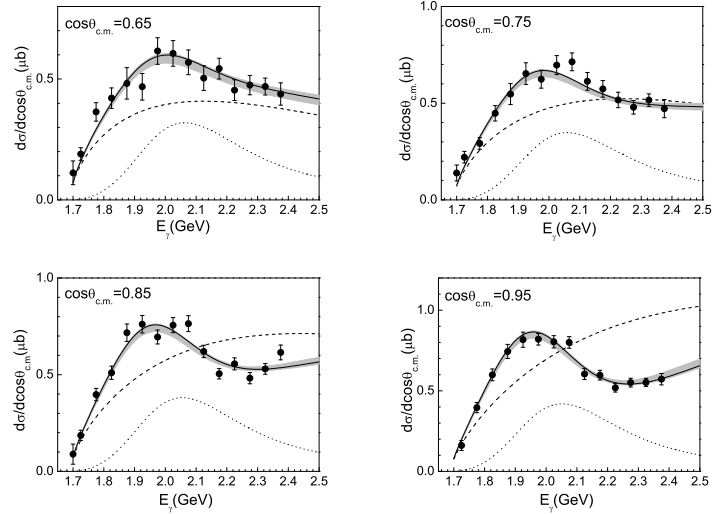}
\caption{$\gamma p \to K^+ \Lambda^*$ differential
$d\sigma/d(\cos\theta_{\rm C.M.})$ cross sections compared with the
LEPS data~\cite{leps}. Dashed and dotted lines show the
contributions from the background and $N^*$ resonance terms,
respectively, while the solid line displays the full result. For
this latter curve we also show the statistical 68\% CL band.}
\end{figure}
\end{center}

With the strong coupling constants obtained from the $\chi^2$ fits,
we can evaluate the branching fraction of the $N^*(2080)$ decay into
the $\Lambda^*K$ channel, which turns out to be $7.5\% \pm 2.8\%$.
We find that the $\Lambda^* K$ mode is quite important. This large
coupling supports the findings of the QM approach~\cite{simonprd58}.

\section{Summary and Conclusions } \label{sec:conclusions}

We have studied the $\gamma p \to \Lambda^*K^+$ reaction at low
energies within a effective Lagrangian approach. In particular, we
have paid an special attention to a bump structure in the
differential cross section at forward $K^+$ angles reported in the
recent SPring-8 LEPS experiment~\cite{leps}. Starting from the
background contributions studied in previous works, we have shown
that this bump might be described thanks to the inclusion of the
nucleon resonance $N^*(2080)$. We have fitted its mass, width and
hadronic $\Lambda^*K^+$ and electromagnetic $ N^* N\gamma $
couplings to data. We have found that this resonance would have a
large decay width into $\Lambda^*K$, which is compatible with the
findings of the QM approach~\cite{simonprd58}.

Other explanations of the observed bump in the SPring-8 LEPS data
are also possible. Indeed, in the very same experimental paper
(Ref.~\cite{leps}) where the data is published, it is suggested that
this structure might be due to a $J^P=\frac{3}{2}^+$ nucleon
resonance, with a mass of $2.11$ GeV and a width of $140$ MeV.
However, a nucleon resonance with these features is not listed in
the PDG book~\cite{pdg2008}. In Ref.~\cite{leps}, it is also
mentioned the possibility of a sizeable contribution from a higher
($J^P=\frac{5}{2}^-$) baryon state and/or the existence of a new
reaction process, for instance, an interference with $\phi$
photo-production~\cite{Hosaka,alvin}.


\begin{theacknowledgments}
 We warmly thank M.J. Vicente--Vacas and J. Martin Camalich for
useful discussions. This work is partly supported by DGI and FEDER
funds, under contract FIS2008-01143/FIS, the Spanish
Ingenio-Consolider 2010 Program CPAN (CSD2007-00042), and
Generalitat Valenciana under contract PROMETEO/2009/0090. We
acknowledge the support of the European Community-Research
Infrastructure Integrating Activity "Study of Strongly Interacting
Matter" (acronym HadronPhysics2, Grant Agreement n. 227431) under
the Seventh Framework Programme of EU. Work supported in part by DFG
(SFB/TR 16, "Subnuclear Structure of Matter"). Ju-Jun Xie
acknowledges Ministerio de Educaci\'on Grant SAB2009-0116.
\end{theacknowledgments}



\bibliographystyle{aipproc}   


\begin{thebibliography}{9}

\bibitem{leps}N.~Muramatsu et al. (LEPS Collaboration), Phys.\ Rev.\ Lett.\ {\bf 103},
012001 (2009); H.~Kohri et al. (LEPS Collaboration), Phys.\ Rev.\
Lett.\ {\bf 104}, 172001 (2010).
%
\bibitem{titovprc7274}A.~I.~Titov, B.~K\"ampfer, S.~Dat\'a, and Y.~Ohashi, Phys.\ Rev.\ C {\bf 72}, 035206 (2005); ibid {\bf C 74},
055206 (2006); A.~Sibirtsev \emph{\textit{et al.}} Eur.\ Phys.\ J.\
A {\bf 31}, 221 (2007).
%
\bibitem{nam}S.~I.~Nam, A.~Hosaka, and H.-C.~Kim, Phys.\ Lett.\ B {\bf 579}, 43 (2004); Phys.\ Rev.\ D {\bf 71}, 114012 (2005): Phys.\
Lett.\ B {\bf 633}, 483 (2006); S.~I.~Nam, Phys.\ Rev.\ C {\bf 81},
015201 (2010).
%
\bibitem{toki}H.~Toki, C.~Garc\'\i a-Recio, and J.~Nieves, Phys.\
Rev.\ D {\bf 77}, 034001 (2008).
%
\bibitem{lamp2} D.~P.~Barber et al. (LAMP2 Collaboration), Z.\ Phys.\, C {\bf 7}, 17 (1980); A.~M.~Boyarski et al., Phys.\ Lett.\ B {\bf 34}, 547 (1971).
%
\bibitem{pdg2008} C.~Amsler \emph{\textit{et al.}} Phys.\ Lett.\ B
\textbf{667}, 1 (2008).
%
\bibitem{simonprd58} S.~Capstick, and W. Roberts, Phys.\ Rev.\ D {\bf 58}, 074011
(1998); S.~Capstick, Phys.\ Rev.\  D {\bf 46}, 2864 (1992).
%
\bibitem{xiejuan}Ju-Jun Xie, and J. Nieves, Phys. Rev. C {\bf 82},
045205 (2010).
%
\bibitem{Hosaka}S.~Ozaki, A.~Hosaka, H.~Nagahiro, and O.~Scholten, Phys.\ Rev.\ C \textbf{80}, 035201
(2009).
%
\bibitem{alvin}A.~Kiswandhi, Ju-Jun~Xie, and S. N. Yang,  Phys.\ Lett.\ B {\bf 691}, 214 (2010).
%

\end{thebibliography}



\end{document}